\documentclass[a4paper,10pt]{article}

\usepackage{graphicx} 
\newcommand{\FSA}{0.5\textwidth}
\newcommand{\FSB}{\textwidth}
\newcommand{\vecr}{\mbox{\boldmath $r$}}

\newcommand{\betwo}{$B(E2, 0^{+}_1 \rightarrow 2^{+}_1)$ }
\newcommand{\twop}{$2^{+}$ }

\begin{document}

\title{
Collective excitations induced by pairing anti-halo effect 
}

\author{Masayuki Yamagami$^{1,2}$\\
{\it \small $^{1}$Heavy Ion Nuclear Physics Laboratory, RIKEN, 
Hirosawa 2-1, Wako, Saitama 351-0198, Japan}\\
{\it \small $^{2}$Department of Physics, Graduate School of Science,
Kyoto University, Kyoto 606-8502, Japan} }

\maketitle

\begin{abstract}
Important features of low-frequency collective vibrational excitations 
in neutron drip line nuclei are studied.
We emphasize that pairing anti-halo effect
in the Hartree-Fock-Bogoliubov (HFB) theory play 
crucial roles to realize collective motions in loosely 
bound nuclei. 
We study the spatial properties of one particle -
one hole (1p-1h) states with/without 
selfconsistent pairing correlations 
by solving simplified HF(B) equations in coordinate space. 
Next, by performing Skyrme-HFB plus 
selfconsistent quasiparticle random phase approximation (QRPA)  
we investigate the first \twop states in neutron rich Ni isotopes.
Three types of calculations, HFB plus QRPA, resonant BCS plus QRPA, 
and RPA are compared.
\end{abstract}


\section{Introduction}

The study of the properties of nuclei far from the $\beta$-stability 
line is one of the most active and challenging 
issues in nuclear structure physics.
Especially interesting phenomena concern the loosely bound neutron
systems close to the neutron drip line.
Low-frequency collective vibrational excitations 
in neutron drip line nuclei are one of the most interesting subjects.
Naively we expect low-frequency vibrational modes
associated with low-density neutron matter (neutron skin and halo). 
A famous one is soft dipole mode \cite{HJ87,Ik92}.  
However it is a fundamental question whether such collective modes 
can really appear or not. 

Vibrational excitations are represented as coherent superposition of 
1p-1h states.
In stable nuclei, because the Fermi energies are deep, 
1p-1h states between tightly bound single-particle states 
having similar spatial characters only contribute, 
and these 1p-1h states concentrate around the nuclear surface.
Around neutron drip line, by contrast, 1p-1h states between
tightly bound , loosely bound, resonance, and non-resonant continuum states,
contribute. 
Because each single-particle state has 
different spatial extent, 1p-1h states among them have different 
spatial characters. 
Therefore it is a non-trivial problem whether vibrational modes
can be realized as a result of coherency between such 1p-1h states.
 
In Section \ref{SEC-phexc} we examine spatial properties of 1p-1h states 
in neutron drip line nuclei.
First we study characters of 1p-1h states without
pairing correlations by solving the Woods-Saxon potential.
Next, by solving simplified HFB equation in coordinate space, 
we emphasize that selfconsistent pairing correlations
play important roles to realize collective motions 
in neutron drip line nuclei.
In Sec.\ref{SEC-QRPA} we perform Skyrme-HFB plus QRPA calculations
for the first $2^{+}$ states in neutron rich Ni isotopes. 
By comparing three types of calculations, HFB plus QRPA, 
resonant BCS plus QRPA, and RPA, 
we show that selfconsistent pairing correlations play important roles 
to realize low-frequency collective vibrational excitations 
in neutron drip line nuclei.


\section{1p-1h states in neutron drip line nuclei} \label{SEC-phexc}


\subsection{1p-1h states without pairing correlations} \label{SUBSEC-NOPAIR}

\begin{figure}[t]
  \begin{center}
    \includegraphics[width=\FSA,clip]{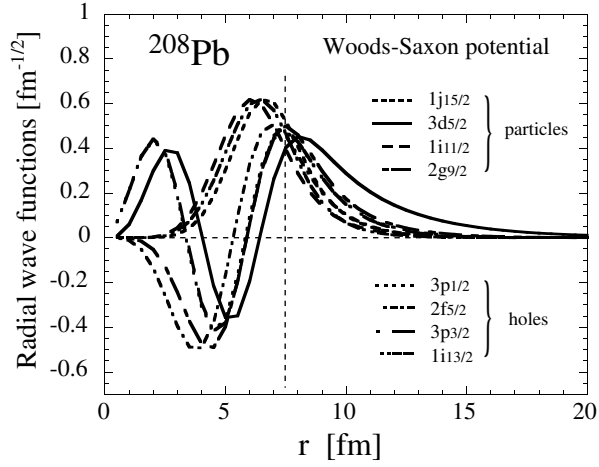}
  \end{center}
  \caption{The neutron radial wave functions around 
Fermi energy in $^{208}$Pb.}
  \label{FIG_Pb_wf}
\end{figure}

\begin{figure}[t]
  \begin{center}
    \includegraphics[width=\FSB,clip]{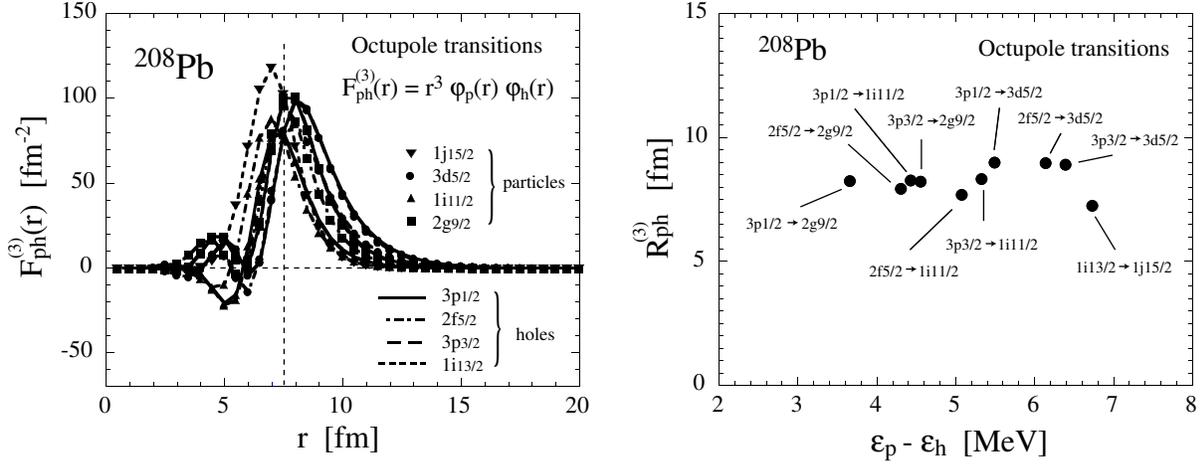}
  \end{center}
  \caption{The spatial distributions of 1p-1h states 
$F_{ph}^{(3)}(r)$ for octupole transitions and 
the corresponding radii $R_{ph}^{(3)}$  in $^{208}$Pb 
are shown as a function of particle-hole energies 
$\varepsilon_p - \varepsilon_h$.}
  \label{FIG_Pb_Fph+Rph}
\end{figure}


We consider the spatial localization of 1p-1h states 
as an important condition for collective vibrational motions.
Before studying 1p-1h states in neutron drip line region, 
we examine neutron 1p-1h states in $^{208}$Pb 
as a typical example of stable nuclei for comparison. 
Because the Fermi energy is about -8 MeV, 
all wave functions around Fermi energy are tightly 
bound and spatially localized as shown in Fig.\ref{FIG_Pb_wf}. 
These wave functions are provided by solving the Woods-Saxon potential.
The remarkable point is that all wave functions have 
large overlaps around the surface region. 
We define the spatial distributions of 1p-1h states as 
\begin{eqnarray}
F_{ph}^{(L)}(r)\equiv \varphi_p (r) \cdot r^L \cdot \varphi_h (r),
\end{eqnarray}
where L is multipolarity of the transition and $\varphi_p$ ($\varphi_h$)
are radial wave functions of particle (hole) states. 
The spatial integration of $F_{ph}^{(L)}(r)$ gives the radial part of 
the particle-hole transition matrix element
$\left<p \left| r^L \right| h \right>
=\int dr F_{ph}^{(L)}(r).$
In Fig.\ref{FIG_Pb_Fph+Rph},  
$F_{ph}^{(L)}(r)$ for octupole transitions are plotted.
The hole (particle) states are indicated by
lines (symbols), and each combination of a line and a symbol
represents a 1p-1h configuration.
As seen, all 1p-1h states are localized around the surface region.
This feature is an important condition for realization of collective motions, 
and this condition is satisfied in stable nuclei. 
For further discussion, we define the "radii" of 1p-1h states $R_{ph}^{(L)}$
as the second moment of $F_{ph}^{(L)}(r)$,
\begin{eqnarray}
R_{ph}^{(L)} \equiv 
\left\{ 
\frac{\int dr r^2 F_{ph}^{(L)} (r) }
{\int dr F_{ph}^{(L)} (r)}
 \right\}^{1/2}.
\end{eqnarray}
In Fig.\ref{FIG_Pb_Fph+Rph}, $R_{ph}^{(3)}$ are plotted 
as a function of particle-hole energies $\varepsilon_p - \varepsilon_h$.
All $R_{ph}^{(3)}$ concentrate around the surface region in stable nuclei.

Next we investigate how the distributions of $R_{ph}^{(L)}$ 
change around neutron drip line region.
We consider neutron rich Ni isotopes as a typical example. 
In the Hartree-Fock (HF) calculation with Skyrme SLy4 parameter, 
$^{86}$Ni is the neutron drip line nucleus.
By taking into account pairing correlations, more neutrons can bound.
However the precise position of the neutron drip line 
depends on the treatment of pairing correlations. 
For example, the drip line nucleus is $^{88}$Ni in Ref.\cite{GS01}
and $^{92}$Ni in Ref.\cite{MD00}. 
The predicted drip line also depends 
on the effective interactions and the frameworks, i.e., relativistic or 
non-relativistic approaches (for example,\cite{MD00,Me98}). 
In the present study we consider up to $^{88}$Ni for further discussion,
because our purpose is to investigate the qualitative aspects of 
vibrational excitations in neutron drip line nuclei. 

\begin{figure}
  \begin{center}
    \includegraphics[width=\FSA,clip]{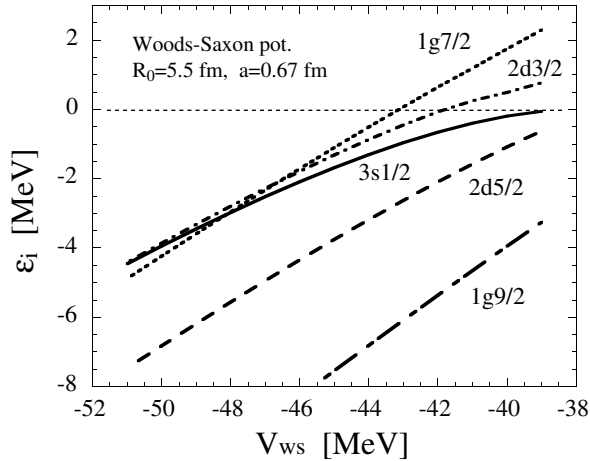}
  \end{center}
  \caption{The neutron single-particle energies around 
$^{86}$Ni are plotted as a function of the depth $V_{WS}$. 
$V_{WS}\approx -41$ MeV corresponds to $^{86}$Ni.}
  \label{FIG_Ni_spe}
\end{figure}

\begin{figure}
  \begin{center}
    \includegraphics[width=\FSB,clip]{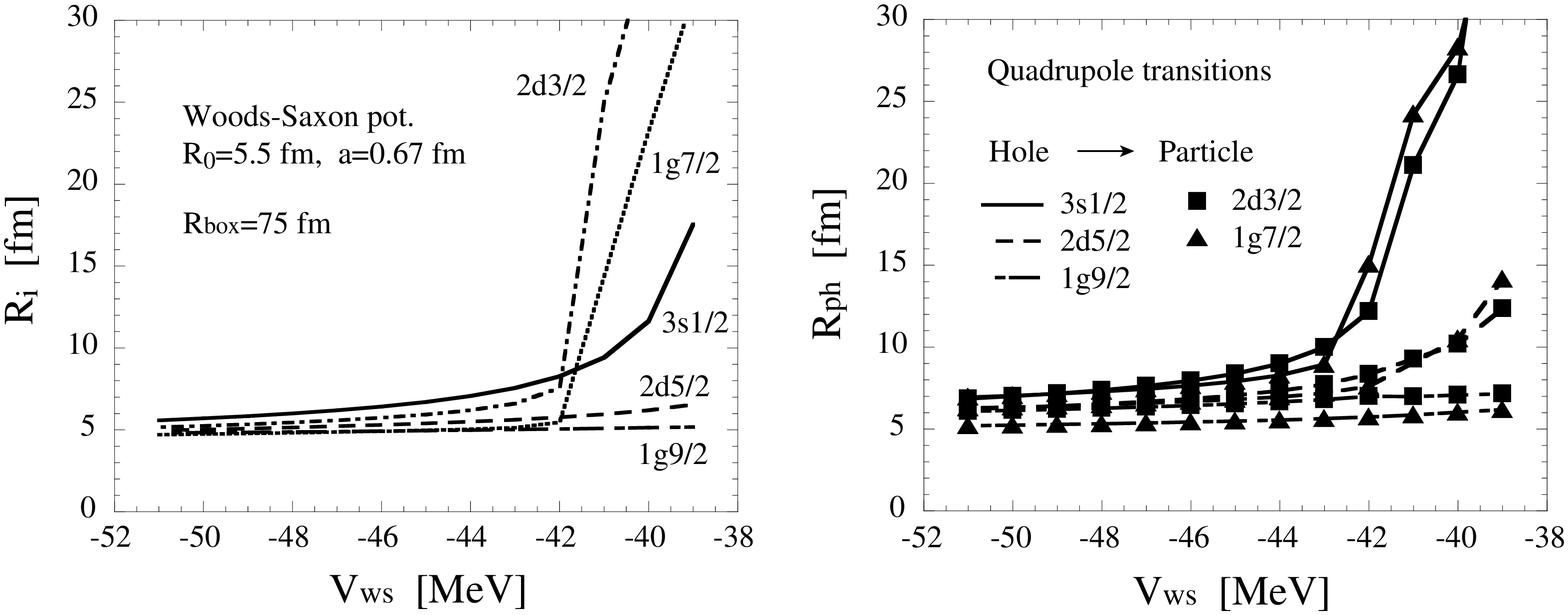}
  \end{center}
  \caption{
The radii of neutron single-particle levels $R_i$ 
and the radii of 1p-1h states $R_{ph}$ 
for quadrupole transitions around 
$^{86}$Ni as a function of the depth $V_{WS}$.}
  \label{FIG_Ni_R+Rph}
\end{figure}

In Fig.\ref{FIG_Ni_spe} the neutron single-particle energies 
around $^{86}$Ni are shown as a function of the depth 
$V_{WS}$ of the Woods-Saxon potential. 
$V_{WS}\approx -41$ MeV corresponds to the single-particle energies
in $^{86}$Ni calculated by HF with SLy4.
We consider 1p-1h states of quadrupole transitions around  
loosely bound $3s_{1/2}$ state (the Fermi level). 
The resonant states, $2d_{3/2}$ and $1g_{7/2}$, are particle states.
In Fig.\ref{FIG_Ni_R+Rph} the radii of single-particle levels 
$R_i$ are shown.
The radii of bound states with large-$\ell$, 
$1g_{9/2}$ and $2d_{5/2}$, don't change as approaching the drip line.
On the other hand, the radius of $3s_{1/2}$  state becomes 
rapidly large in the limit of zero binding energy, i.e., neutron halo.
Concerning particle states, $2d_{3/2}$ and $1g_{7/2}$, 
while the single-particle energies are negative, these radii change 
slowly as a function of $V_{WS}$. On the other hand, once the 
single-particle energies become positive (resonant states), 
the radii increase suddenly.   
In Fig.\ref{FIG_Ni_R+Rph} the radii of 1p-1h states $R_{ph}$ 
for quadrupole transitions are plotted as a function of $V_{WS}$.
The hole (particle) states are indicated by lines (symbols).
Up to $V_{WS} \leq -44$ MeV, corresponding to stable nuclei region, 
all $R_{ph}$ are localized around the surface region.
By contrast, in the drip line region $V_{WS} \geq -44$ MeV, 
$R_{ph}$ strongly depend on the particle-hole configurations.
Size of $R_{ph}$ is mainly determined by the spatial extent of 
hole wave functions. 
The localization of 1p-1h states becomes more sparse and 
the condition for realization of collective motions is not satisfied
as approaching the drip line.
We may conclude that neutron drip line region is under the situation 
that collective motions are suppressed and single-particle like 
excitations appear dominantly. 
In next subsection we discuss how
this situation is significantly modified by including selfconsistent
pairing correlations.  

\subsection{Roles of selfconsistent pairing correlations} \label{SUBSEC-PAIR}


It is a well-known fact that pairing correlations are important 
ingredient to describe collective vibrational excitations.
In this study we want to emphasize that selfconsistent pairing correlations 
described in the HFB theory play important roles in neutron drip line nuclei
in non-trivial ways.

Pairing anti-halo effect \cite{BD00} is an important property 
of the HFB theory, and qualitatively modifies the asymptotic behavior of 
the quasiparticle wave functions.
In HF the asymptotic behavior of the bound states wave functions 
($\varepsilon_{lj,n}<0$) for $r \rightarrow \infty$ is
\begin{eqnarray}
\varphi _{lj}^{HF} \left( {\varepsilon _{lj,n} ,r} \right) 
\to \exp \left( { - \alpha _{lj,n} r} \right),
\end{eqnarray}
where
$
\alpha _{lj,n}  = \sqrt { -2m \varepsilon_{lj,n} } / \hbar. 
$
The decay constant $\alpha _{lj,n}$ can approach zero 
for $\varepsilon_{lj,n} \rightarrow 0$. 
For positive energy states ($\varepsilon >0$), 
\begin{eqnarray}
\varphi _{lj}^{HF} \left( {\varepsilon ,r} \right) \to C_{lj} \left[ {\cos \left( {\delta _{lj} } \right)rj_l \left( {\beta r} \right) - \sin \left( {\delta _{lj} } \right)rn_l \left( {\beta r} \right)} \right]
\end{eqnarray}
where
$
\beta  = \sqrt{ 2m \varepsilon } / \hbar.
$
$j_l$ and $n_l$ are spherical Bessel and Neumann functions, 
and $\delta_{lj}$ is the phase shift corresponding to the angular momentum
$(lj)$.

By taking into account pairing correlations, 
irrespective of selfconsistent or non-selfconsistent pairing,
quasiparticle wave functions have two components, 
hole wave functions $v_{lj} (r)$ and
particle wave functions $u_{lj} (r)$.
In HF plus BCS calculations both $v_{lj} (r)$ and $u_{lj} (r)$ 
are just proportional to HF single-particle wave functions,
\begin{eqnarray}
\begin{array}{l}
 u_{lj}^{} \left( {\varepsilon _{lj,n} ,r} \right) = 
u_{lj,n} \,\varphi _{lj}^{HF} \left( {\varepsilon _{lj,n} ,r} \right) \\ 
 v_{lj}^{} \left( {\varepsilon _{lj,n} ,r} \right) = 
v_{lj,n} \,\varphi _{lj}^{HF} \left( {\varepsilon _{lj,n} ,r} \right), \\ 
 \end{array}
\label{EQ-BCS-ASY}
\end{eqnarray}
where $v_{lj,n}$ ($u_{lj,n}$) are occupation (unoccupation) amplitudes.
Therefore both hole and particle wave functions 
have the same asymptotic behavior for $r \rightarrow \infty$.
Eqs.(\ref{EQ-BCS-ASY}) are usually defined only for bound states. 
Recently the HF plus resonant BCS method
was proposed \cite{SL97,SG00}. In this approach
resonant states are take into account for particle states.
Applicability of the resonant BCS method for ground state properties 
is discussed in Ref.\cite{GS01}.
In HFB, by contrast, the asymptotic behavior of the quasiparticle 
wave functions is qualitatively different.
We examine the HFB equation in coordinate space \cite{Bu80,DF84,BS87},
\begin{eqnarray}
\left( {\begin{array}{*{20}c}
   {h_{lj} \left( r \right) - \lambda } & {\Delta \left( r \right)}  \\
   {\Delta \left( r \right)} & { - h_{lj} \left( r \right) + \lambda }  \\
\end{array}} \right)\left( {\begin{array}{*{20}c}
   {u_{lj} \left(E_{qp}, r \right)}  \\
   {v_{lj} \left(E_{qp}, r \right)}  \\
\end{array}} \right) = E_{qp} \left( {\begin{array}{*{20}c}
   {u_{lj} \left(E_{qp}, r \right)}  \\
   {v_{lj} \left(E_{qp}, r \right)}  \\
\end{array}} \right), \label{EQ-HFB}
\end{eqnarray}
where $h_{lj}(r)$ and $\Delta_{lj}(r)$ are the mean field 
and the pairing field respectively. 
For the discrete region $(E_{qp}+\lambda)<0$, 
both hole and particle wave functions decay exponentially
at infinity,  
\begin{eqnarray}
\begin{array}{l}
 u_{lj} \left( E_{lj,n}, r \right) \to \exp \left( { - \alpha _{lj,n}^{(1)} r} \right) \\
 v_{lj} \left( E_{lj,n}, r \right) \to \exp \left( { - \beta _{lj,n}^{(1)} r} \right) \\ 
 \end{array}
\end{eqnarray}
where
$
\alpha _{lj,n}^{(1)}  = \sqrt { -2m \left(E_{lj,n}+\lambda\right)}/\hbar
$ 
and
$
\beta _{lj,n}^{(1)}  = \sqrt { 2m \left(E_{lj,n}  - \lambda\right)}/\hbar.
$
For the continuum region $(E_{qp}+\lambda)>0$, 
\begin{eqnarray}
\begin{array}{l}
 u_{lj} \left( {E,r} \right) \to C_{lj} \left[ {\cos \left( {\delta _{lj} } \right)rj_l \left( {\alpha _{}^{(2)} r} \right) - \sin \left( {\delta _{lj} } \right)rn_l \left( {\alpha _{}^{(2)} r} \right)} \right] \\ 
 v_{lj} \left( {E,r} \right) \to \exp \left( { - \beta ^{(2)} r} \right) \\ 
 \end{array}
\end{eqnarray}
where
$
\alpha^{(2)}  = \sqrt { 2m \left(E + \lambda \right)} /\hbar 
$
and
$
 \beta^{(2)}  = \sqrt { 2m \left(E - \lambda \right)} /\hbar.
$
The hole wave functions always decay exponentially.
The minimum value of the decay constant $\beta$ is estimated as following.
If a discrete quasiparticle state with 
minimum energy $E_{\mu}$ is present, the minimum value of 
the decay constant becomes 
$\beta_{\mu}^{(1)}=\sqrt { 2m \left(E_{\mu}-\lambda\right)}/\hbar$.
The HFB quasiparticle energies $E_{\mu}$ are well approximated by the 
canonical quasiparticle energies \cite{DN96}, 
$E_{\mu}\simeq E_{\mu}^{can} \equiv 
\sqrt{(\epsilon_{\mu}^{can}-\lambda)^2 +(\Delta_{\mu}^{can})^2}$, 
where $\epsilon_{\mu}^{can}$ and $\Delta_{\mu}^{can}$ are the canonical 
single-particle energy and the canonical pairing gap.
This means that $\beta_{\mu}^{(1)}$ stays at finite value
\begin{eqnarray}
\beta_{\mu}^{(1)}\rightarrow \sqrt { 2m E_{\mu}}/\hbar
\geq \sqrt { 2m \Delta_{\mu}^{can}}/\hbar >0
\end{eqnarray} 
with finite $\Delta_{\mu}^{can}$ for $\lambda \rightarrow 0$.
Selfconsistent pairing correlations affect the asymptotic behavior of 
all hole wave functions, and
especially act against a development of an infinite 
rms radius that characterizes the wave functions of $s$ and $p$ states 
in the limit of vanishing binding energy.
We call this localization of hole wave functions 
"pairing anti-halo effect".
Thought this terminology was used for the 
localization of the normal density distribution $\rho(r)$ 
in Ref.\cite{BD00}, the microscopic mechanism is the same.
In neutron drip line region where discrete HFB solutions don't exist,
the analysis is more complicated. 
However, as numerically shown in Ref.\cite{BD00}, 
pairing anti-halo effect plays the roles to localize hole
wave functions in similar way.

\begin{figure}
  \begin{center}
    \includegraphics[width=\FSB,clip]{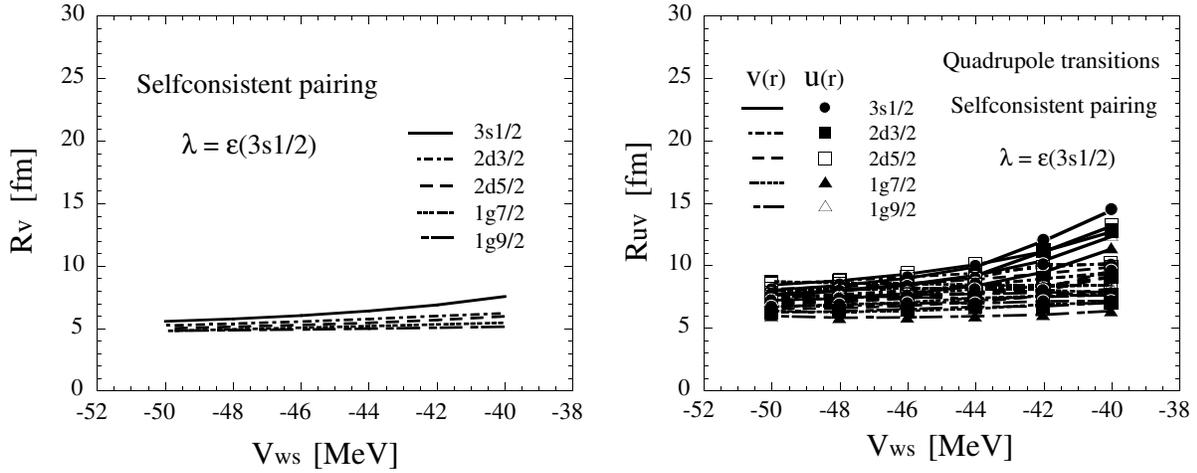}
  \end{center}
  \caption{The radii of hole wave functions $R_v$ 
and the radii of 1p-1h states $R_{uv}$ in HFB 
around $^{86}$Ni as a function of the depth $V_{WS}$.}
  \label{FIG_HFB_Rv+Rvu}
\end{figure}

We solve the coordinate space HFB equation Eq.(\ref{EQ-HFB})
with box boundary conditions $R_{box}=30$ fm.
For the HF potential the Woods-Saxon potential is used.
The pairing potential is selfconsistently derived 
by the density dependent pairing interaction,
\begin{eqnarray}
V_{pair} (\vecr,\vecr') = V_{pair} 
\left[ 1 - \frac{\rho \left( \vecr \right)}{\rho_c} \right]
\delta \left( \vecr - \vecr' \right). \label{EQ-DDPI}
\end{eqnarray}
The parameters are $V_{pair}=-350$ MeV fm$^{-3}$ and
$\rho_c=0.16/2$ fm$^{-3}$ with the cut-off energy $E_{cut}=50$ MeV.
$\rho_c$ is taken to be the half of the saturation density, 
because Eq.(\ref{EQ-HFB}) is solved only for neutrons. 
This pairing parameter set gives the averaged pairing 
gap $\bar{\Delta}\approx 1.5$ MeV in the region $-51$ MeV $< V_{WS} < -39$
MeV.
The average pairing gap is defined as the integral of the pairing field
with the abnormal density $\tilde{\rho}(\vec{r})$, 
\begin{eqnarray}
\bar{\Delta} = \frac{ \int d\vec{r} \tilde{\rho} (\vec{r}) \Delta(\vec{r}) }
{\int d\vec{r} \tilde{\rho} (\vec{r})}. 
\end{eqnarray}
In Fig.\ref{FIG_HFB_Rv+Rvu}, the radii of hole 
wave functions around $^{86}$Ni provided by HFB calculations 
are plotted.
By comparing the radii of the single-particle states without 
selfconsistent pairing correlations (HF and HF plus resonant BCS
in Fig.\ref{FIG_Ni_R+Rph}), the spatial distributions of the hole 
wave functions are completely different.
Due to pairing anti-halo effect, the radii in HFB 
spatially concentrate around the surface region.


We consider how spatial characters of 1p-1h states 
are modified by taking into account selfconsistent pairing 
correlations. 
First of all, by including pairing correlations, irrespective of 
selfconsistent or non-selfconsistent, the possible 1p-1h configurations 
increase considerably, i.e., 
not only particle-hole but also particle-particle 
and hole-hole configurations can contribute.
We define a spatial distribution of 1p-1h state
from a hole part $v_{lj}(E)$ to a particle part $u_{l'j'}(E')$, 
\begin{eqnarray}
F^{(L)} [u_{l'j'}(E'),v_{lj}(E)](r) \equiv \ u_{l'j'}(E',r) 
\cdot r^{L} \cdot v_{lj}(E,r),
\end{eqnarray}
and the radius,
\begin{eqnarray}
R^{(L)} [u_{l'j'}(E'),v_{lj}(E)] \equiv \left\{
\frac{\int dr \ r^2 F^{(L)} [u_{l'j'}(E'),v_{lj}(E)](r)}
{\int dr \ F^{(L)} [u_{l'j'}(E'),v_{lj}(E)](r)}\right\}^{1/2}.
\end{eqnarray}
We use shorthand notations
$F_{uv}^{(L)}(r)$ and $R_{uv}^{(L)}$, 
unless these don't cause confusions.
In Fig.\ref{FIG_HFB_Rv+Rvu} the radii $R_{uv}^{(2)}$ in HFB are plotted.
As we have mentioned in subsec.\ref{SUBSEC-NOPAIR}, 
the radii of 1p-1h states are mainly determined by 
the spatial extent of hole states.
Because the radii $R_{v}$ in HFB are localized
due to the pairing anti-halo effect, 
the radii $R_{uv}$ in HFB concentrate around the surface region.
This localization of $R_{uv}$ can happen by treating pairing 
correlation selfconsistently.
In resonant BCS, such localization can not appear.
Generally speaking, the spatial distribution of 
particle-particle configurations between resonant states are 
always far outside from the surface region in resonant BCS.  

\section{HFB plus QRPA calculations} \label{SEC-QRPA}


\subsection{Formulation}

We consider the first \twop states in neutron rich 
Ni isotopes by Skyrme-HFB plus selfconsistent 
QRPA calculations. 
By selfconsistent we mean that the HFB mean fields are 
determined selfconsistently from an effective force 
and the residual interaction of the QRPA problem is derived from the same
force. The QRPA problem is solved by the response function method 
in coordinate space. A detailed account of the method can be found in
Ref.\cite{YG04,YK03,KS02}. 
Here, we just recall the main steps of the calculation.
The QRPA Green's function ${\bf G}$ is a solution of a Bethe-Salpeter equation,
\begin{equation}
{\bf G} = {\bf G}_0 +{\bf G}_0 {\bf V}{\bf G}~.
\end{equation}
The knowledge of ${\bf G}$ allows one to construct the response function of
the system to a general external field, and the strength distribution of the
transition operator corresponding to the chosen field is just proportional
to the imaginary part of the response function.
The residual interaction ${\bf V}$ between quasiparticles is derived from the 
Hamiltonian density $\left<H\right>$ of Skyrme interaction by the so-called Landau procedure,
\begin{equation}
V_{\alpha \beta} = 
\frac{\partial ^2 \left<H\right> }
{
\partial \rho_\beta  \partial \rho_{\bar \alpha }
}. \label{eqLM}
\end{equation}
The index $\alpha(\alpha=1,2,3)$ stands for particle-hole ($ph$), 
particle-particle ($pp$), and hole-hole ($hh$) channels.
The notation ${\bar \alpha }$ means that 
whenever $\alpha$ is $pp$ ($hh$) then ${\bar \alpha }$ is $hh$ ($pp$).
The residual interaction ${\bf V}$ has an explicit momentum dependence, 
\begin{equation}
{\bf V}(\vecr,\vecr')={\bf F}[\overleftarrow{\Delta}_U +\overleftarrow{\Delta}_V, 
\overrightarrow{\Delta}_U +\overrightarrow{\Delta}_V,
\overleftarrow{\nabla}_U \pm \overleftarrow{\nabla}_V,
\overrightarrow{\nabla}_U \pm \overrightarrow{\nabla}_V]
\delta (\vecr-\vecr'). \label{EQ_VRESI}
\end{equation}
The explicit form of the form factor ${\bf F}$ is shown in Ref.\cite{YG04}.
These momentum dependence are explicitly treated in our calculation.
Because we calculate only natural parity (non spin-flip) excitations, 
we drop the spin-spin part of the residual interaction. 
The Coulomb and spin-orbit residual interactions are also dropped.

The ground states are given by Skyrme-HFB calculations.
The HFB equation is diagonalized on a Skyrme-HF basis calculated 
in coordinate space with box boundary conditions $R_{box}=20$ fm. 
Spherical symmetry is imposed on quasiparticle wave functions.
The quasiparticle cut-off energy is taken to be   
$E_{cut}=50$ MeV, and the angular momentum cut-off is $\ell_{max}=7\hbar$ 
in our HFB and QRPA calculations.
The Skyrme parameter SLy4 \cite{CB98}
are used for the HF mean-field, and the density-dependent,
zero-range pairing interaction Eq.(\ref{EQ-DDPI})
is adopted for the pairing field.  
The parameters are $V_{pair}=-555$ MeV fm$^{-3}$, $\rho_c=0.16$ fm$^{-3}$, 
so as to give the average neutron pairing gap 
$\bar{\Delta}_n \approx 12/\sqrt{A}$ in $^{86}$Ni.


\subsection{Ni isotopes}

\begin{figure}[tb]
  \begin{center}
    \includegraphics[width=\FSB,clip]{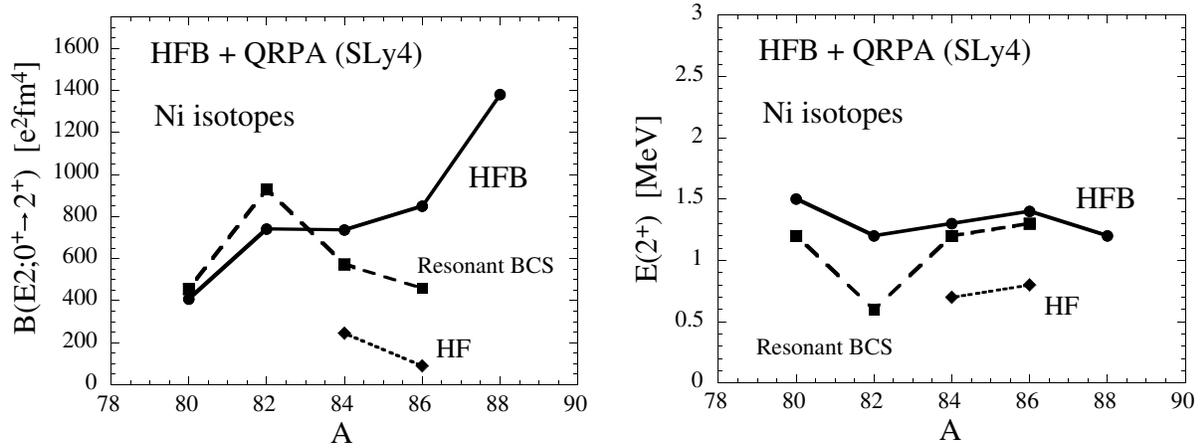}
  \end{center}
  \caption{The \betwo values and the excitation energies 
of the first \twop states in neutron rich Ni isotopes 
calculated by HFB plus selfconsistent QRPA, 
resonant BCS plus QRPA, and RPA with Skyrme SLy4 force.}
  \label{FIG_QRPA_E2}
\end{figure}

In order to emphasize the importance of pairing anti-halo effect 
for  realization of collective vibrational excitations, 
we compare three types of calculations, HFB plus QRPA, 
resonant BCS plus QRPA (no pairing anti-halo effect), 
and RPA (no pairing) for the first \twop states in neutron rich Ni 
isotopes.
In Fig.\ref{FIG_QRPA_E2} the \betwo values
and excitation energies of the first \twop states 
in neutron rich Ni isotopes up to $^{88}$Ni in HFB plus QRAP, 
and $^{86}$Ni for resonant BCS plus QRPA corresponding
the neutron drip line are shown.
To construct the quasiparticle wave functions without 
pairing anti-halo effect, we perform resonant BCS calculations.
In resonant BCS calculations for the ground states,
we include resonant states (localized states) for particle states.
To identify the resonant states in calculations 
with box boundary conditions $R_{box}=20$ fm, 
we introduced an additional cut-off radius
$R_{BCS}=10$ fm and the single-particle orbits having 
the radii less than $R_{BCS}$ are taken into account 
for the pairing problems of the ground states. 
The resonant $1g_{7/2}$ and $2d_{3/2}$ states are included. 
The orbits with the radii larger than $R_{BCS}$
are included only for QRPA model space.
Because we intend qualitative discussion 
of microscopic aspects of vibrational excitations,
we performed constant-$\Delta$
BCS calculations to avoid technical complexities
that are inherent in non-selfconsistent calculations
(lack of selfconsistency in the pairing potential,
different energy cut-off scheme for HF plus 
resonant BCS calculations $E_{cut}=5$ MeV
and QRPA calculations $E_{cut}=50$ MeV, and 
neglecting the contribution of non-resonant continuum state 
that plays important roles in pairing correlations 
in neutron drip line nuclei \cite{BD00,BD99,HM03}).  
The constant pairing gap $\Delta=0.8$ MeV is chosen to 
reproduce low-lying quasiparticle energies in Skyrme-HFB calculations.
We have checked the sensitivity of the pairing gap parameter,
and the calculated excitation energies and 
the $B(E2)$ values are qualitatively the same.
For the residual interaction we used the exactly same interaction
in HFB plus QRPA calculations.

From $^{80}$Ni to $^{82}$Ni, HFB plus QRPA and resonant BCS plus QRPA
calculations give qualitatively similar results.
The energies decrease and the $B(E2)$ values increase.
On the other hand, as approaching the neutron drip line, the calculated 
$B(E2)$ values exhibit different behavior.
In HFB plus QRPA calculations the $B(E2)$ values increase monotonically.
On the other hand, the $B(E2)$ values suddenly decrease from  
$^{82}$Ni to $^{86}$Ni in resonant BCS plus QRPA.
The main qualitative difference between HFB plus QRPA 
and resonant BCS plus QRPA is the presence of pairing anti-halo effect.
In resonant BCS, due to lack of pairing anti-halo effect, 
particle-particle configurations between resonant states, 
$1g_{7/2}$ and $2d_{3/2}$, and also particle-hole configurations 
with hole $3s_{1/2}$ state are spatially decoupled to 
the other 1p-1h states, 
and can not participate in the collective excitations.
The results of HF plus RPA calculations for $^{84}$Ni and $^{86}$Ni
are also shown.
These excitations have single-particle like characters, 
and the excitation energies are almost the same with specific
1p-1h energies ($2d_{5/2} \rightarrow 3s_{1/2}$ in $^{84}$Ni 
and $3s_{1/2} \rightarrow 2d_{3/2}$ in $^{86}$Ni).   
The $B(E2)$ values in resonant BCS plus QRPA are larger than 
the values in HF plus RPA, because hole-hole configurations can contribute
to make the collective modes in resonant BCS plus QRPA calculations.

For another example that indicates the importance of selfconsistent 
pairing correlations,
we have shown that neutron pairing correlations 
play important roles to explain the experimentally observed 
anomalously large $B(E2)$ values in neutron rich nuclei, 
$^{32}$Mg and $^{30}$Ne \cite{YG04}.
This is an actual example that pairing anti-halo effect 
plays important roles to realize collective motions 
in neutron drip line region.


\section{Conclusion}  \label{SEC-CONC}

We discussed the important microscopic aspects of 
low-frequency collective vibrational excitations 
in neutron drip line nuclei.
We emphasized that pairing anti-halo effect in the HFB theory plays 
important roles to realize collective motions in loosely bound nuclei. 
We performed Skyrme-HFB plus selfconsistent QRPA calculations  
for the first \twop states in neutron rich Ni isotopes.
By comparing three types of calculations, HFB plus QRPA, 
resonant BCS plus QRPA, and RPA, 
the importance of pairing anti-halo effect was shown in 
realistic calculations.


\section{ACKNOWLEDGMENTS}

We acknowledge 
Professor K. Matsuyanagi for valuable discussions.
Numerical computation in this work was carried out at the 
Yukawa Institute Computer Facility.

%
%

\newpage

%
%





\end{document}